\documentstyle[epsf,prd,aps,textcomp,floats,multicol]{revtex}
\newcommand{\be}{\begin{equation}}
\newcommand{\ee}{\end{equation}}
\newcommand {\Ell}{{\cal L}}
\newcommand {\bea}{\begin{eqnarray}}
\newcommand {\eea}{\end{eqnarray}}

\def\expec#1{\langle#1\rangle}
\def\eq#1{{\frenchspacing eq.}~(\ref{#1})}
\def\etal{{\frenchspacing\it et al.}}
\def\ie{{\frenchspacing\it i.e.}}
\def\eg{{\frenchspacing\it e.g.}}
\def\etc{{\frenchspacing\it etc.}}

\def\Mpc{{\rm Mpc}}
\def\l{\ell}

\def\C{{\bf C}}
\def\E{{\bf E}}
\def\I{{\bf I}}
\def\X{{\bf X}}
\def\d{{\bf d}}
\def\e{\mathbf\varepsilon}
\def\p{{\bf p}}
\def\q{{\bf q}}
\def\x{{\bf x}}
\def\y{{\bf y}}
\def\z{{\bf z}}

\begin{document}
%\twocolumn[\hsize\textwidth\columnwidth\hsize\csname@twocolumnfalse\endcsname
%\title{Growth of inhomogeneities in a generalized Chaplygin gas Universe}
%\title{CMBfit: Extreme compression of the WMAP likelihood surface}
\title{CMBfit: Rapid WMAP likelihood calculations with normal
parameters}
\author{H\aa vard B. Sandvik$^1$, Max Tegmark$^1$, Xiaomin Wang$^1$ and Matias Zaldarriaga$^2$}
\address{$^1$Dept. of Physics, Univ. of Pennsylvania, Philadelphia, PA 19104, USA; sandvik@hep.upenn.edu \\
$^2$Dept. of Physics, Harvard University, Cambridge, MA 02138,
USA} \maketitle

\begin{abstract}
We present a method for ultra-fast confrontation of the WMAP
cosmic microwave background observations with theoretical models,
implemented as a publicly available software package called CMBfit,
useful for anyone wishing to measure cosmological parameters by combining WMAP with
other observations.
The method takes advantage of the underlying physics by transforming
into a set of parameters where the WMAP likelihood surface is
accurately fit by the exponential of a quartic or sextic polynomial.
Building on previous physics based approximations by
Hu {\etal}, Kosowsky {\etal} and Chu {\etal},
it combines their speed with precision cosmology grade accuracy.
A Fortran code for computing the WMAP likelihood for a given set of parameters is
provided, pre-calibrated against CMBfast, accurate to
$\Delta\ln\Ell\sim 0.05$ over the entire $2\sigma$ region of the
parameter space for 6 parameter ``vanilla'' $\Lambda CDM$ models.
We also provide 7-parameter fits including spatial curvature, gravitational waves
and a running spectral index.
\end{abstract}
%]
\begin{multicols}{2}
\section{Introduction}
The impressive sensitivity of the
long awaited Wilkinson Microwave
Anisotropy Probe (WMAP) data
allows for unprecedented constraints on cosmological models
\cite{bennett03,hinshaw03,kogut03,page03,spergel03,verde03}.
%In the months since the release the data has sparked an enormous interest and
%though not unexpected interest\cite{afterWMAP}.
%The
%perhaps most spectacular result is the almost complete lack of
%surprises. In fact t
The measurements have strengthened the case for the cosmological
concordance model\cite{verde03}, the inflationary $\Lambda$CDM
model, a flat Universe which is currently accelerating due to
mysterious dark energy, where most matter is in the form of
collisionless cold dark matter, and where the initial conditions
for the fluctuations are adiabatic. Because of well known
cosmic microwave background (CMB) parameter
degeneracies \cite{hu00,spergel03,bridle},
%It is
%well known that the CMB does not alone give all these constraints,
%well known parameter degeneracies in the cosmic microwave
%background (CMB) prevent that\cite{hu00}.
the true impressiveness
of the data is most clearly demonstrated by either imposing reasonable
priors, combining the data with complimentary data sets, or both.
The most recent precision data-sets are the Sloan Digital Sky Survey (SDSS)
galaxy power spectrum
(SDSS) \cite{SDSS} and a new Supernovae Ia compilation \cite{tonry03},
and combining these with the WMAP constraints have further narrowed error
bars \cite{SDSS2}, giving us cosmological parameters at a precision not
thought possible only few years ago.
%Indeed it can now be argued
%that cosmology has ceased to be the playground for theorists that
%it once was (although the amount of energy that has been put into
%the field of non-trivial topologies since the WMAP release may
%indicate otherwise).

%indeed a vast number of prospective cosmological scenarios may now
%be discarded or at the very least strongly constrained by use of
%CMB data and reasonable priors \cite{spergel03}. Even stronger
%constraints may be obtained by combining WMAP with other
%cosmological surveys (eg. the Sloan Digital Sky Survey galaxy
%power spectrum\cite{SDSS}).

So how are CMB data-sets used to estimate parameters? Although
constraints on cosmological parameters from the CMB can in
principle be extracted directly from the maps themselves, this is
effectively prohibited by the huge computational power necessary
to perform the required likelihood calculations (although for a
novel approach, see \cite{wandelt03}). Instead, the more efficient
parameter estimation method commonly employed uses the angular
power spectrum of the map as an intermediate
step\cite{tegmark97,bond87,bond98}.

Likelihoods on the basis of power spectrum constraints are much
faster to calculate, the slowest step being the computation of the
angular power spectrum numerically. This can be done either through
integration of the full Boltzmann equation (with CMBfast \cite{CMBfast} or the
modifications
% MT: Matias keeps reminding me that CAMB is build on CMBfast, not a fully independent code
CAMB\cite{camb} or CMBEASY\cite{cmbeasy}) or using an approximate
shortcut such as DASh\cite{dash}. Although steady improvements
in both computer power and algorithm performance have made these
calculations significantly faster, the CMB power spectrum
likelihood calculation is still the bottleneck in any parameter
estimation process by many orders of magnitude.

Another problem with the estimation process are the
near-degeneracies between some cosmological parameters. Elongated,
banana shaped contours on the likelihood surface make search
algorithms less efficient. Several authors have advocated various
transformations from cosmological parameters to ``physical"
parameters more directly linked to features in the power
spectrum\cite{hu00,kosowsky02}. Chu {\etal} \cite{chu02} advocated
a new such set of ``normal" parameters whose probability
distributions are well fitted by Gaussian distributions.

Our key idea explored in this paper is that if the likelihood
function is roughly a multivariate Gaussian in the transformed
parameters, then it should be very accurately approximated by the
exponential of a convenient higher-order polynomial. A Gaussian
likelihood surface $\Ell$ corresponds to the log-likelihood
$\ln\Ell$ being quadratic, and a quadratic Taylor expansion is of
course an accurate approximation of any function near its maximum.
Very far from the maximum, the Gaussian approximation again
becomes accurate, since both it and the true likelihood $\Ell$ are
vanishingly small. We will see that in most cases, small cubic and
quartic corrections to $\ln\Ell$ help provide a near-perfect fit
to $\Ell$ everywhere. The WMAP team employed such quartic fits to
determine reasonable step sizes for their Markov Chain Monte Carlo
search algorithm \cite{verde03}. Here we go further and show how
polynomial fits can conveniently store essentially all the
cosmological information inherent in the WMAP power spectra. This
reduces calculation of CMB likelihoods to simply evaluating the
exponential of an $n^{th}$ order polynomial. Although CMBfast
still needs to be run in order to obtain a sufficient sampling of
the likelihood surface, this only has to be done once for each
model space and dataset. For the models explored in this paper, it
has already been done and the polynomial coefficients have been
obtained. The polynomial fit can thus be used to run Monte Carlo
Markov Chains (MCMC) including various non-CMB data-sets  many
orders of magnitude faster, dramatically reducing the need for
computing power for cosmological parameter estimation purposes
involving WMAP data. This is an improvement on importance sampling
methods which notoriously depletes the chain of points as the
added data sets shift or narrow the confidence regions. It means
joint likelihood analyses between WMAP and other surveys which
would previously take weeks or months of computer time on a good
workstation being finished in an afternoon.

As mentioned, the last few years have seen the rise of a
cosmological concordance model\cite{xiaomin02,spergel03,SDSS2}, the $\Lambda$CDM
flat inflationary cosmology which has been confirmed and
strengthened by WMAP, SDSS and new supernova observations. However
as cosmological ideas and trends change, we choose to keep an open
mind and allow for extensions to the 6-parameter inflationary $\Lambda$CDM
model by including models with spatial curvature, gravitational waves and a
running scalar spectral index.
%, non-negligible
%neutrino contributions as well as different values of the dark
%energy equation of state $w$.
To allow the user to include CMB polarization information separately,
we also perform separate likelihood fits excluding and including WMAP polarization
data for the 6 parameter scenario.

\section{Method}\label{METHOD}

Our approach in this paper consists of three steps:
\begin{enumerate}
\item{Acquire a sample of the likelihood surface as a function of
cosmological parameters. This is done through a Markov Chain Monte
Carlo sampling of parameter space.}
\item{Transform from
cosmological parameter space into normal parameters. This will
make the likelihood surface close to Gaussian in these parameters,
thereby increasing significantly the accuracy of the polynomial
fit.}
%\item{Transforming the ``normal" parameters into their eigenbasis
%in order to numerically stabilize the fit}
\item{Fit of the log-likelihood surface to an $n^{th}$ order
polynomial. The polynomial degree $n$ is optimized to the
the likelihood-surface sampling density using a training set/test set
approach.}
\end{enumerate}
Below we describe each of the above
steps in detail.
\subsection{The Cosmological Parameters}
We follow standard work\cite{spergel03,xiaomin02} in the field and
parameterize our cosmological model with the $13$ parameters
\begin{equation}
{\mathbf p} = \left(\tau, \Omega_k, \Omega_\Lambda, w, \omega_d,
\omega_b, f_{\nu}, A_s, n_s, \alpha,r,n_t,b\right) \label{params}
\end{equation}
These parameters are the reionization optical depth, $\tau$,
curvature energy density $\Omega_k$, the dark energy density
$\Omega_\Lambda$, the dark energy equation of state $w$, the
physical dark matter and baryon densities $\omega_{d}\equiv
\Omega_d h^2$ and $\omega_b \equiv \Omega_b h^2$, the fraction of
dark matter that is warm (massive neutrinos) $f_\nu$, the
primordial amplitudes and tilts of the scalar and tensor
fluctuations respectively $A_s,n_s,A_t,n_t$ and the running of the
scalar tilt $\alpha$. Here $A_s, n_s$ and $\alpha$ are defined by
the \emph{Ansatz} $P_*(k) = A_s (k/k_{eval})^{n_s + \alpha \ln
k}$, and similarly for the tensor case. $b$ is the galaxy bias
$b^2 = P_{galaxy}(k)/P(k)$. For a comprehensive cosmological parameter summary,
see Table 1 of \cite{SDSS2}.

As stated in the introduction, we base our work around the
adiabatic $\Lambda$CDM cosmological model,
$\{\tau,\Omega_\Lambda,\omega_d,\omega_b, n_s, A_s\}$, a 6-parameter
subspace of the 13 parameters. This is close to the minimal number of free parameters
needed explain the data (the one exception being $n_s$, which is still consistent with unity)
and assumes a pure cosmological
constant and negligible spatial curvature, tensor fluctuations, running tilt,
warm dark matter or hot dark matter
As \cite{spergel03,SDSS2}, we also consider models with
added ``spice" such as curvature, tensor contributions and
running tilt. We  confine ourselves to a maximum of 7
parameters per model, \ie, to models with $\Lambda$CDM + a 7$^{th}$ free parameter.

%The basic six parameters are the reionization optical depth,
%$\tau$, the dark energy density $\Omega_\Lambda$, physical dark
%matter and baryon densities $\omega_{d}\equiv \Omega_d h^2$ and
%$\omega_b \equiv \Omega_b h^2$, and the amplitude and tilt of the
%scalar fluctuation primordial power spectrum $A_s$ and $n_s$.

%The latter two are defined from the Ansatz for the primordial power spectrum,
%\begin{equation}
%    P_*(k) = A_s (k / k_s)^{n_s + \alpha \ln k},
%\end{equation}
%where $n_s = 1$ represents a scale invariant spectrum.

%Similar to \cite{SDSS} we go further and ``spice up" this vanilla
%$\Lambda$CDM model by considering models with added curvature
%$\Omega_k = 1 - \Omega_{tot}$, running tilt $\alpha$, tensor $r$
%perturbations or a physical density of massive neutrinos
%$\omega_{\nu} \equiv \Omega_\nu h^2$.

\subsection{The Likelihood}

The fundamental quantity that one wants to estimate is the probability distribution function
(PDF) of the parameters vector $\p$, $P(\p|\d)$, given the data, $\d$,
and whatever prior assumptions and knowledge we may have about the
parameters. The quantity we directly evaluate, however, is the
probability of measuring the data given the parameters,
$P(\d|\p)$ through a goodness-of-fit test. It is this
distribution, when thought of as a function of the parameters, that
we refer to as the \emph{likelihood}, $\Ell(\p)\equiv P(\d|\p)$.
The probability distribution function for the parameters is then
related to the likelihood through Bayes' theorem:
\be\label{Peq}
    P(\p|\d) \propto P(\d|\p)
    P_{prior}(\p)
\ee
where $P_{prior}$ is the prior probability distribution of the
parameters.

For ideal, full-sky, noiseless experiments, exact likelihood
calculation is simple and fast. However, due to foreground contamination
(the Galaxy, point sources etc.), only a fraction of the sky can be
used for analysis. This leads to correlations between different
multipoles and it becomes computationally prohibitive to calculate
the exact likelihood function. Consequently, various approximations
exist on which much work has been focused\cite{verde03,bond} (for
an excellent review of CMB likelihood calculations see
\cite{jaffe03}).
%
%in fact in all our likelihood calculations we employ the
%likelihood calculator routines supplied by the WMAP team.
%
%Naively the likelihood function can be approximated by $\Ell =
%\exp(-\chi^2/2)$, $\chi^2$ being a simple chi-square goodness of
%fit of a theoretical power spectrum to the estimated $C_l's$.
%
%In reality the $C_l's$ are correlated due to cut-sky maps and
%other effects such as non-Gaussian uncertainties in the bandpowers
%so caution must be taken to calculate $\chi^2$\cite{bond}.
In all our WMAP likelihood calculations we employ the latest version of
the likelihood
approximation routines supplied by the WMAP team. These routines take
all effects into account, use an optimal combination of the
various approximations, and are well tested through
simulations\cite{verde03}. As input, they take the CMB power spectrum, which
we compute with CMBfast.

\subsection{Fitting to an $n^{th}$ order polynomial}
\label{fitting}

Let us now look at how the polynomial fit is performed in detail.
We start with a sample of N points $\p_1,...,\p_N$ in the
$d$-dimensional parameter space where the likelihood $\Ell(\p)$
has been evaluated, and wish to fit $\ln\Ell(\p)$ to a polynomial.
Let $\expec{\p}$ denote the average of the parameter vectors
$\p_i$ in our sample and let
$\C\equiv\expec{\p\p^t}-\expec{\p}\expec{\p}^t$ denote the
corresponding covariance matrix. To improve the numerical
stability of our fitting, we work with transformed parameters that
have zero mean and unit covariance matrix\footnote{The advantage
of diagonalising the parameter covariance matrix was also pointed
out by \cite{chu02}}:
\be
\z\equiv\E(\p-\expec{\p}),
\ee
where $\E$ is a matrix satisfying $\E\C\E^t=\I$ so that
$\expec{\z\z^t}=\E\expec{(\p-\expec{\p})(\p-\expec{\p})^t}\E^t=\I$, the identity matrix.
There are many such choices of $\E$ --- we make the choice where
the rows of $\E$ are the
eigenvectors the parameter covariance matrix $\C$ divided by the
corresponding eigenvalues, so our transformed parameters $\z$ can be interpreted as
simply uncorrelated eigenparameters rescaled
so that they all vary on the same scale (with unit variance).
This transformation turns out to be crucial, changing the matrix-inversion
below from quite ill-conditioned to numerically well-conditioned.

A $n^{th}$ order polynomial in these $d$ transformed parameters
has $M = {n+d \choose n}={(n+d)!\over n!d!}$ terms:
\bea
y\equiv \log{\Ell} &=& q_0 + \sum_i q^i_1 z_i +
\sum_{i_1\le i_2} q_2^{i_1i_2}z_{i_1} z_{i_2} +
%\sum{i_1\le i_2 \le i_3} q_3^{i_1i_2i_3}z_{i_1} z_{i_2} z_{i_3}
\nonumber \\& &
% +
%\sum{i\le j\le k \le l} q_4^{ijkl}z_i z_j z_k z_l +\cdots,\\& &
 + \sum_{i_1\le i_2 \le \cdots \le i_d} q_n^{i_1i_2\cdots
i_d}z_{i_1}z_{i_2}\cdots z_{i_d} \label{quartic}
\eea
%\bea
%y\equiv \log{\Ell} &=& q_0 + \Sigma_i q^i_1 \delta_i +
%\Sigma_{i\le j}q_2^{ij}\delta_i \delta_j + \Sigma_{i\le j \le k}
%q_3^{ijk}\delta_i \delta_j \delta_k \nonumber\\& &
% +
%\sum{i\le j\le k
%\le l} q_4^{ijkl}\delta_i \delta_j \delta_k \delta_l + \cdots,\\
%\label{quartic}
%\eea
%where $\delta_i = (\alpha_i - \mu_i^0)/\sigma_i$, and $\mu_i^0$ is
%the mean value of the parameter.
We assemble all necessary products of $z_i$'s into an
$M$-dimensional vector
\be
\x = \{1,z_1,\cdots,z_d,z_1 z_1,\cdots,\Pi_{i=1,n}z_i\},
\ee
and the corresponding coefficients $q$ into another $M$-dimensional vector
\be
\q =
\{q_0,q_1^1,\cdots,q_1^d,q_1^{11},\cdots,q_n^{d...d},\cdots\},
\ee
which simplifies eq.~(\ref{quartic}) to
\be
    y = \x\cdot\q.
\ee
We now assemble the $N$ measured log-likelihoods $y_i$ from our Monte Carlo Markov Chain
into an $N$-dimensional vector $\y$ and the corresponding $\x$-vectors into
an $N \times M$-dimensional matrix $\X$, so
the $N$-dimensional vector of residual errors $\e$ from our polynominal
fit is
\be
 \e\equiv\y-\X\q.
\ee
We choose the fit that minimizes the rms residual.
\ie, that minimizes $|\e|^2$. Differentiating with respect to $\q$ gives
the standard least-squares result
\be
    \q = \left(\X^t\X\right)^{-1}\X^t\y.
\ee
Thus the minimizing the sum of squares in the
end comes down to the inversion of an $M\times M$ matrix. The
size of the matrix, $M=(n+d)!/n!d!$,
depends on both number of parameters and the
polynomial degree, ranging from $M = 210$ for a 6 parameter 4th order fit to
$M = 1716$ for a 7 parameter 6th order fit (see
table \ref{ncoeffs} for the number of coefficients for various
relevant cases).

\begin{table*}[ht]
\begin{tabular}{|c|ccccc|}
Parameters & $n=2$ & $n=3$ & $n=4$ & $n=5$ & $n=6$ \\
\hline 5 & 21 & 56 & 126 & 252 & 462 \\
6  &  28 & 84 & 210 & 462 & 924 \\
7 &  36 & 120 & 330 & 792 & 1716 \\
8  &  45 & 165 & 495 & 1287 & 3003 \\
9  &  55 & 220 & 715 & 2002 & 5005 \\
10  &  66 & 286 & 1001 & 3003 & 8008 \\
11  &  78 & 364 & 1365 & 4368 & 12376 \\ \hline
\end{tabular}
\caption{Number of coefficients for a model with $d$ parameters
fitting to an $n^{th}$ order polynomial. The number of coefficients
range from 28 for a 6 parameter $2^{nd}$ order fit to a whopping
12376 for an 11 dimensional model and $6^{th}$ order polynomial.
%As matrix inversion is a $N^3$ process, it is clear that for some
%cases this will be the limiting factor rather than sample size.
% MT: No, 122376 is still  a piece of cake!
}
\label{ncoeffs}
\end{table*}

Chu et.al.\cite{chu02} have illuminated the problems of fitting a
Gaussian directly to the 6- (or higher) dimensional likelihood
surfaces and have argued that the surfaces may be too sparsely
sampled in these dimensions. Consequently \cite{chu02} fits to the
2D marginalized distributions and reconstructs the 6D likelihood
function from this. %We take the alternative view
Our interpretation is
%It is worth
%noting that we thus take a more optimistic view than the one
%expressed by \cite{chu02}, who claim the likelihood surface is too
%sparsely sampled in 6 (or higher) dimensions to fit to. This is
%claimed despite using only a Gaussian fit, which amounts to 12
%coefficients for the 6 parameter model. We instead take the view
that the difficulties with fitting a quadratic polynomial to
the 6 or 7 dimensional log-likelihood surface shows that the
likelihood surface deviates too much from a Gaussian and that a
higher order polynomial is required to reproduce the likelihood to
sufficient accuracy. This interpretation is shared by \cite{verde03} who
use a $4^{th}$ order polynomial to calculate MCMC step sizes.

It is an advantage of our approach that we do \emph{not} rely
strictly on the chain we fit to being a fair statistical sample of
the likelihood. Indeed we only need the \emph{value} of the
likelihood at a sufficient number of points, and we are as such
insensitive to statistical errors such as sampling errors and poor
mixing. The way that the input points $\p_i$ sample parameter
space tells the fitting algorithm how important we consider
residuals in various places. Since our points are distributed as
the WMAP likelihood itself, the fit will be accurate in those
parts of parameter space that are consistent with the data. If our
fits are combined with complementary data sets, high accuracy is
of course only necessary in the small jointly allowed region of
parameter space, and this accuracy can optionally be further
improved by including non-CMB data to determine how to sample the
CMB likelihood surface.

Clearly the sample size (the number of steps in the input Monte
Carlo Markov Chain) along with the dimensionality of the parameter
space determines how densely the likelihood surface is sampled. In
order to make a best possible fit for the 7 parameter models we
therefore include the points from the 6 parameter chain in the
fit, thus placing extra statistical weight on the vanilla
parameter substance. This allows us to use the higher parameter
fit to get excellent results for the 6 parameter case as well in
addition to reducing polynomial artifacts.

%\subsubsection{Accuracy}
Of course the polynomial has complete freedom outside the sampled
region, which means that for degree $n>2$ the fit will generally blow up in
regions far from the origin. This means that once a search
algorithm ventures outside the allowed region, it may find
unphysical areas of huge likelihoods, much higher than the real
maximum. We find that this artifact is efficiently eliminated by
replacing the polynomial fit by a Gaussian
$y=e^{-r^2/2}$  outside some large radius
$r\equiv |\z| = (z_1^2 + \cdots + z_d^2)^{1/2}$ in the transformed parameter space.\

To ensure that we do not introduce significant polynomial
artifacts within the sampled region, we use a standard training set/test set approach.
We run the fit on, \eg, 70\% of the chain and
test the fit on the remainder of the chain.
As the polynomial degree is increased,
the training errors will inevitably get smaller since
there are more degrees of freedom, while the polynomial eventually
develops unphysical small-scale wiggles in between sample points. This problem
is quantified by measuring the errors in the test set, allowing us to
identify the optimal polynomial degree as the point where
the test set error is minimized. In the limit of very large sample size $N$, the test and
training errors approach the same value for any given polynomial degree $n$.

%In the limit of infinite chains the test errors will approach the
%value of the training errors, however in reality comparing the two
%values indicates the goodness-of-fit and identifies an optimal
%polynomial degree as the point when the test error is minimized.

%We will perform the fit for the full WMAP TT + TE power spectra
%for 6 parameter $\Lambda$CDM and various 7 parameter models, as
%well as a separate fit for the WMAP TT power spectrum alone for
%the 6 parameter case.

\subsection{Markov Chain Monte Carlo}
To make an accurate polynomial fit, we need a sufficiently large
sample of the likelihood surface as a function of the cosmological
parameters. This is done by Markov Chain Monte Carlo (MCMC)
sampling of the parameter space through the use of the
Metropolis-Hastings algorithm
\cite{mackay,cosmomc,knox,christensen,Metropolis,Hastings,Gilks96,GelmanRubin92,Christensen01,Slosar03}.
When implemented correctly, this is a very effective method for
explorations of parameter space, and we briefly review the concept
here. What we want is to generate samples $\p_i$, $i=$0, 1, 2, ...
from the probability distribution $P(\p)$ of \eq{Peq}. The method
consists of the following steps:
\begin{enumerate}
\item
We start by choosing a starting point $\p_0$ in
parameter space, and evaluate the corresponding value of the
probability distribution $P(\p_0)$.
\item Next we draw a candidate point
$\p_{i+1}$ from a \emph{proposal} density $Q(\p_{i+1}|\p_i)$ and
calculate the ratio
\be
    a = { Q(\p_i|\p_{i+1}) \over Q(\p_{i+1}|\p_i)} {P(\p_{i+1}) \over P(\p_i)}.
\ee
\item
If $a \ge 1 $ we accept this new point, add the new point to the
chain, and repeat the process starting from the new point. If $a < 1$, we
accept the new state with probability $a$, otherwise we reject it,
write the current state to the chain \emph{again} and make another
draw from the proposal density $Q$.
\end{enumerate}
After an initial burn-in period which depends heavily on the
initial position in parameter space (the length of this burn-in
can be as short as 100 steps whilst some chains still have not
converged after several thousand steps), the chain starts sampling
randomly from the distribution $P(x)$, allowing for calculation of
all relevant statistics such as means and variances of the
parameters. The choice of proposal density is of great importance
for the algorithm performance as it defines a characteristic step
size in parameter space. Too small a value and the chain will
exhibit poor mixing, an excessive step size and the chain will
converge very slowly since almost all candidate points get rejected.
The acceptance ratio is a common measure of how successful a chain is.
However, whereas a low acceptance ratio certainly
demonstrates poor performance, high acceptance ratio can be an artifact of
too small step size, which makes successive points in the chain highly correlated.
As discussed in \cite{SDSS2}, a better figure of merit is the
chain \emph{correlation length}, as it determines the effective number of
independent points in the chain.
%We therefore use this test to ensure optimal performance.

Our particular MCMC implementation is described in Appendix A of \cite{SDSS2},
and gradually optimizes the proposal density $Q(\p'|\p)$ using the
data itself. Once this learning phase is complete
(typically after about 2000 steps, which are then discarded),
our proposal density $Q(\p'|\p)$ is a multivariate Gaussian
in the jump vector $\p'-\p$
with the same covariance matrix as the sample points $\p_i$ themselves.
This guarantees optimum performance of the Metropolis algorithm by
minimizing the number of jumps outside high confidence regions,
whilst still ensuring good mixing.
A very similar eigenbasis approach to jumping
has been successfully used in other recent MCMC codes, notably
\cite{chu02,cosmomc,ZahnZalda03}.

For each case described in the next section we generate several chains,
with different initial conditions, which we pass through a number
of convergence and mixing tests given in \cite{verde03} and \cite{mackay}.

\subsection{CMB observables and normal parameters}
The issue of what we can actually \emph{measure} from the CMB is
of fundamental importance, since it helps to clarify which
constraints come directly from the CMB and which constraints can
only be found by combining CMB with other cosmological probes.
This has been studied in detail by Hu {\etal} \cite{hu00}, who
suggested that much of the then available information in the temperature power
spectrum could in fact be compressed into only four observables, the
overall horizontal
position of the first peak plus 3 peak height ratios. This work was
re-visited by the WMAP team\cite{page03} and
others\cite{doran1,doran2} have similarly studied the effects of the
parameters, including dark energy, on CMB peak locations and
spacings. From such studies one can understand the degeneracies
between cosmological parameters, by studying their effect on these
quantities.\footnote{This work pre-dates WMAP and other recent CMB
surveys. The enormous improvement in the data will by itself have
reduced the degeneracies to some degree, adding more information
to the power spectrum.}

Kosowsky {\etal} \cite{kosowsky02} go further along these lines
and propose a set of ``physical" parameters to which the power
spectrum $C_\l$s have an approximately linear response. This allows for fast
calculation of power spectra around the fiducial model. The
approach was taken further by
\cite{chu02} in realising that a linear response to these
parameters in the $C_\l$'s should result in the logarithm of the
likelihood function being well represented by a 2nd order
polynomial, \ie, the likelihood function should be close to
Gaussian in these parameters. This approach resulted in another,
similar set of parameters dubbed ``normal'', since they had an approximately
normal distribution.

%\subsection{Our choice of normal parameters}
In this work, we use a best of all worlds approach and employ a
core set of normal parameters which are a combination of the
choices available in the above mentioned literature, with some
improvements. We use a core set of 6 parameters corresponding to
the flat $\Lambda$CDM model. Specifically,
$\{\tau,\Omega_\Lambda,\Omega_d h^2, \Omega_b h^2, n_s, A_s \}$
cast into $\{e^{-2\tau},\Theta^E_s, h_3, h_2, t, A_p\}$. These new
parameters are the physical damping due to the optical depth, an
analytic fit to the angle subtended by the acoustic scale, the
1st-to-3rd peak ratio, the 1st-to-2nd peak ratio with tilt
dependence removed, the physical effect of $n_s$ (\eg, tilt), and
the fluctuation amplitude at the WMAP pivot point. We will now go
through them in detail one by one.

\subsubsection{The Acoustic Scale Parameter, $\Theta_s$}

The comoving angular diameter distance at decoupling is
given by
\be
    D_A(a_{dec}) = {c \over H_0} \int_{a_{dec}}^1 { dx \over \sqrt{ \Omega_k x^2 +
    \Omega_{de} x^{(1-3w)} + \Omega_m x }} \label{angdiamdist}
\ee
where we have ignored radiation density since the moment of
interest, decoupling, is well within matter domination. Note that
the integrand is a function of only \emph{three} parameters,
$\Omega_k, \Omega_{de}$ and $w$. If we assume that the scale
factor at decoupling is constant, the integral is also dependent upon
only these three parameters.

Although a numerical evaluation of the above integral is trivial,
it is not as fast as we would wish and would quickly become the
dominant obstacle in the polynomial likelihood calculation. There
are several reasonably good analytic fitting formulae out
there\cite{hu00}. However none of them are accurate enough for
our needs, so we also perform a polynomial fit for $D_A$. We
rewrite the expression for the angular diameter distance as
\be
D_A(a) \approx D_A(a)^E = {c \over H_0} {2 \over \sqrt{\Omega_m}}
\times d(\Omega_k,\Omega_\Lambda, w)\label{daanal}
\ee
%where $d$ can be approximated by $d=d(\Omega_k, \Omega_\Lambda,
%w)$ which is an analytic approximation to the integral and equals
where $d$ is an analytic approximation to the integral and equals 1
for a matter dominated universe. We factor out $\sqrt{\Omega_m}
= \sqrt{1-\Omega_k - \Omega_\Lambda}$ to remove a troublesome
inverse square root which is difficult to fit with a polynomial
expansion. We fit $d(\Omega_k,\Omega_\Lambda, w)$ with a $5^{th}$
order polynomial expansion,
% MT: It's NOT a Taylor expansion unless the coeffs are given by derivatives at the origin
done by calculating $d$ numerically for
several hundred thousand points in $(\Omega_k,\Omega_\Lambda,w)$-space
and fitting to this sample. Our main errors then come
from the assumption of a constant recombination redshift; however
this fit is good to the $\sim 0.1 \% $ level, and performs notably
better than the fit of \cite{hu00} for non-flat ($\Omega_k \neq 0$)
and dynamical dark energy ($w \neq -1$) scenarios. This fit can be downloaded
as part of our publicly available CMBfit software package.

The comoving sound horizon at decoupling is defined as
\begin{equation}
r_s = \int_0^{t_{dec}} { c_s(t) dt \over a(t)}
\end{equation}
where $c_s(t)$ is the sound speed for the baryon-photon fluid at
time $t$, well approximated by
\begin{equation}
c_s^2 = {1 \over 3} (1+3\rho_b / 4\rho_\gamma)^{-1}.
\end{equation}
Using the relation $dt/a = da/(a^2H)$ and the Friedman equation, we
can write this similarly to eqn.(\ref{angdiamdist})
%\begin{multicols}{2}
\begin{equation}
r_s(a_{dec}) = {1 \over H_0 \sqrt{3}} \int_0^{a_{dec}} {
(1+{3\Omega_b\over 4 \Omega_\gamma}x)^{-1/2} dx \over [(\Omega_k
x^2+\Omega_{de} x^{1-3w} +\Omega_m x + \Omega_r)]^{1/2}}
\end{equation}
If we assume that vacuum energy can be ignored at last scattering,
this can be readily integrated to give
\begin{equation}
r_s = { 2\sqrt{3} \over 3H_0\sqrt{\Omega_m}} \left(
R_{*}(1+z_*)\right)^{-1/2} \ln {
    \sqrt{1+R_*}+\sqrt{R_*+r_*R_{*}} \over 1+\sqrt{r_*R_{*}}}
    \label{rsanal}
\end{equation}
where the photon-baryon and radiation-matter ratios at last
scattering are given by\cite{hu95,hu00}
\bea
    R_* & \equiv & {3 \rho_b(z_*) \over 4
    \rho_{\gamma}(z_*)}=30 \omega_b(z_*/10^3)^{-1} \\
    r_* & \equiv & {\rho_r(z_*) \over \rho_m(z_*)} = 0.042 \omega_m^{-1}(z_*/10^3),
\eea
with a redshift of last scattering
\bea
    z_* &\approx& 1048\left(1+0.00124 \omega_b^{-0.738} \right)
    \left(1+g_1 \omega_m^{g_2}\right) \\
    g_1 &=& 0.0783\omega_b^{-0.238}\left(1+39.5\omega_b^{0.763}\right)^{-1}
    \\
    g_2 &=& 0.560 \left(1 + 21.1 \omega_b^{1.81}\right)^{-1}.
\eea

%Hu et.al. provides an analytic fit for the sound horizon at early
%times\cite{hu95,kosowsky02}
%\begin{equation}
%    r_s = {2 \sqrt{3} \over 3}(\Omega_0 H_0)^{-1/2} \left({a_{eq}
%    \over R_{eq}}\right)^{1/2} \times \ln {
%    \sqrt{1+R}+\sqrt{R+R_{eq}} \over 1+\sqrt{R_{eq}}}
%\end{equation}
% Where $R = 3\rho_b / 4\rho_\gamma$ and is proportional to
%$a(t)$. $R_{eq}= R(t_{eq})$ and $t_{eq}$ is the time of
%radiation-matter equality.

The sound horizon and the angular diameter distance at the time of
decoupling combine to give the angle subtended by the sound
horizon at last scattering. In degrees, it is given by
\begin{equation}
\Theta_s \equiv { r_s(a_{dec}) \over D_A(a_{dec})} {180 \over\pi}.
\end{equation}
This has been verified to be an excellent choice for a normal
parameter \cite{kosowsky02,chu02} and is indeed one of the best
constrained cosmological parameters \cite{spergel03,SDSS}. We use
this parameter, except replacing the exact integrals
$D_A(a_{dec})$ and $r_s(a_{dec})$ with their analytic
approximations, equations~(\ref{daanal}) and (\ref{rsanal}).
%;$\Theta_s^E \approx \Theta_s$.

%\end{multicols}

\subsubsection{The peak ratios $h_2$, $h_3$ and the scalar tilt parameter $t$}
Hu {\etal} \cite{hu00} (and the recent re-analysis of\cite{page03})
define parameter fits to the ratios of the 2nd and 3rd peaks to
the first peak. Again these are near ideal normal parameters since
they are directly measurable from the power spectrum. However
there is one problem. Our requirement is a parameter set with
which to replace the cosmological parameters. $H_3$ is mostly
dependent on $\Omega_m h^2$ and $H_2$ most heavily dependent on
$\Omega_b h^2$ --- however,  they both also depend on the tilt. In other
words, three cosmological parameters come together to form two
observables. Thus to obtain a corresponding three-parameter set,
we factor out the tilt-dependence from $H_2$ (Page {\etal}
\cite{page03}) and $H_3$ (Hu {\etal} \cite{hu00})and create the
variable set $\{h_2, h_3, t\}$, where
%\begin{equation}
\bea
h_2 &= &H_2/2.42^{n_s - 1} \nonumber \\
 &=& 0.0264 \omega_b^{-0.762}
\times e^{-0.476\left(ln(25.5\omega_b+1.84\omega_m)\right)^2}
\eea
%\end{equation}
and
\begin{eqnarray}
h_3 &=& H_3/3.6^{n_s -1} = 2.17 \left(1+(\omega_b
/0.044)^2\right)^{-1}\times \nonumber \\ & & \omega_m^{0.59}
\left(1+1.63(1-\omega_b/0.071) \right)^{-1}.
\end{eqnarray}
The parameter $t$ is given by a slight modification of the formula
used by \cite{chu02} in order to minimize correlation with
$\omega_b$
\begin{equation}
t = \left({\omega_b \over 0.024}\right)^{-0.5233} 2^{n_s-1}
\end{equation}

\subsubsection{The amplitude at the pivot point}
For the amplitude we again use the choice of \cite{chu02}. This
choice removes the near perfect degeneracy with the opacity
$e^{-2\tau}$ due to a non-zero optical depth. It also evaluates
the amplitude at a more optimal pivot point rather than at the
arbitrary $k=0.05\Mpc^{-1}$, such as to remove degeneracy with the
tilt $n_s$. However we make two modifications: We change the
choice of pivot-point, as this is dependent on the data set and
needs to be updated to optimize results for WMAP. We also remove a
strong correlation with $\omega_m$ empirically. The resulting
formula is
\begin{equation}
    A^* = A_s e^{-2\tau}
    \left({k \over k_{pivot}}\right)^{n_s -1} \omega_m^{-0.568},
\end{equation}
where $k_{pivot} = 0.041\Mpc^{-1}$ (this choice minimizes $\Delta A^* / A^*$
using WMAP temperature and polarization information; the corresponding optimal value is $k_{pivot}
= 0.037\Mpc^{-1}$ using WMAP temperature information alone).

%\subsubsection{The physical dark energy density $\omega_Q$}
\subsubsection{The non-vanilla parameters, $\omega_\Lambda$, $\alpha$ and $r$}

For the 7 parameter case where the assumption of spatial flatness
is relaxed, the choice of an extra normal parameter is not
obvious. Since we use $\Theta_s$ as one of our parameters, in
principle any of $\{\Omega_\Lambda, \Omega_k, h\}$ could be used.
Since there is now an extra free component in the Friedman
equation, we choose instead to go with the \emph{physical} dark
energy density $\omega_Q \equiv \Omega_Q h^2$. This has most of
the desirable properties we  seek, and consequently it gives very
small errors in the polynomial fit. We could in fact equally well
have chosen the \emph{physical} curvature density $\omega_k \equiv
\Omega_k h^2$ as this gives very similar results. Both perform
significantly better than any of the above three.

%\subsubsection{The neutrino fraction $f_{\nu}$}
%The CMB is well known for not constraining the neutrino
%contributions very significantly. It constrains well the physical
%(warm + cold) dark matter and baryon densities $\omega_{dm} \equiv
%\Omega_{dm} h^2$ and $\omega_{b} \equiv \Omega_{b} h^2$ however,
%so in order to find a ``loosely" Gaussian parameter we choose the
%fraction of dark matter that is warm, $f_{\nu} \equiv \omega_{\nu}
%/ \omega{dm}$. We have shown elsewhere\cite{SDSS} that indeed this
%is not a parameter well constrained by WMAP, but which is strongly
%constrained by galaxy surveys.

%\subsubsection{The running of the tilt $\alpha$}
The running of the tilt is defined as $\alpha = d n / d \ln k$.
When taken as a free 7th parameter, it has a distribution function
which is very nearly Gaussian (seen in fig
\ref{wmappollicia7paralfig}). Thus we use this parameter directly
as a normal parameter.

%\subsubsection{The tensor to scalar ratio $r$}
For the tensor contribution, we use as our normal parameter
the tensor to scalar ratio, $r\equiv A_t / A_s$,
where $A_t$ and $A_s$ are defined earlier as
the CMBfast tensor and scalar fluctuation amplitudes respectively,
evaluated at $k=0.05\Mpc^{-1}$.

\section{RESULTS}\label{RESULTS}
In this section, we present the results of the fits. We estimate
the errors in the fitting and in particular display the
marginalized likelihoods obtained by running Markov Chain Monte
Carlo chains using the our fits in place of CMBfast.

%\begin{widetext}
\subsection{Fitting accuracy}
In principle a data set may be fitted to any accuracy by using a
polynomial of sufficiently high order. However, this will introduce
unphysical polynomial artifacts, which will ruin the method's
applicability. As explained in section \ref{METHOD}, we therefore
split our data into a training set and a test set. This allows
us to identify the optimal order of the polynomial that we fit to.

\begin{figure*}[ht]
\centerline{\epsfxsize=9.0cm\epsffile{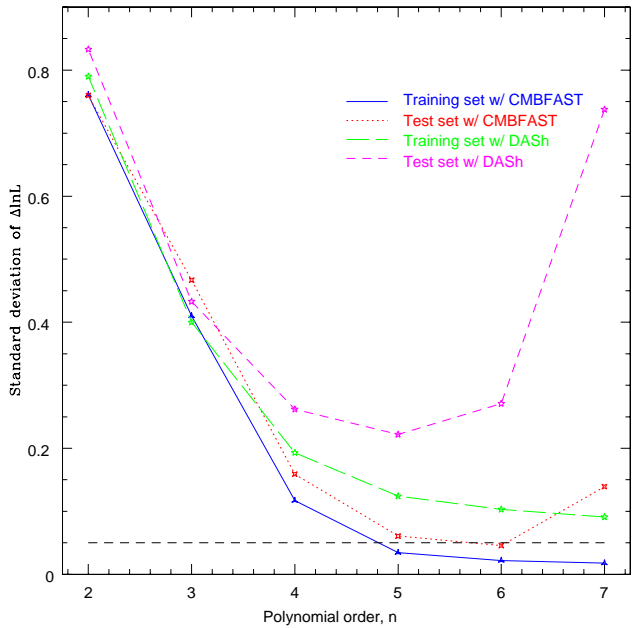}}
\caption{R.m.s.~fitting error $\Delta\ln\Ell$
% We also call this quantity epsilon above
% \equiv \ln\Ell_{fit} - \ln \Ell$
for the 6 parameter $\Lambda$CDM case.
The plot compares the training and test sets
for fits based on chains using CMBfast and chains using DASh. For
the particular chain-length used for the CMBfast case, we see that
a 6th degree fit is optimal.
% LET'S NOT IRRITATE LLOYD: (WE ALREADY COMMENT ON IT BELOW)
%Note also the inherent
%difficulty in fitting the DASh calculated likelihoods by a
%polynomial, which is due to more random scatter in these
%likelihoods compared to CMBfast.
} \label{testtrain}
\end{figure*}

This approach is illustrated for the 6 parameter $\Lambda$CDM case
in figure \ref{testtrain}, where we plot the fitting accuracy for
2nd through 7th order polynomials, showing the difference between test
and training sets. The training set errors predictably fall with
increasing polynomial degree, as there are more degrees of freedom with
which to fit the data. The test set errors, however, have a clear
minimum (in this case for $n = 6$) which is what we choose as
optimal polynomial fit. This optimal polynomial degree depends strongly on
how long a chain is used for the fit. A large sample allows us to
go to higher order, whereas for a small sample, even 3rd or
4th order polynomials may be ill fated.

Due to theoretical prejudice for
the probability distributions to be close to Gaussian as well as
relatively smooth, it is possible that some of the errors in
fitting to the CMBfast-calculated likelihoods could be due to
inaccuracies in CMBfast itself rather than our polynomial approximation.
although we have not attempted to test or quantify this, this may be worth exploring
in future work.

Figure \ref{testtrain} also shows that we obtained slightly larger
random scatter when computing power spectra with
DASh \cite{dash} instead of CMBfast.
This does not to appear to be a limitation of DASh itself,
since the accuracy is greatly improved when using the
latest version of DASh with transfer functions from the latest
version of CMBfast (Knox 2003, private communication).
From here on, we use CMBfast % 4.4
% MT: THAT'S NOT TRUE: LICIA USED 4.2 and I USED  4.3, NO?
to calculate all our likelihood samples.

% Figure \ref{testtrain} also shows that we encountered difficulties
% when fitting polynomials to likelihoods calculated using DASh \cite{dash}.
% This indicates the presence of a slight random scatter.
% This scatter can be traced back to us using
% CMBfast 3.2 to compute the DASh grid rather than the more recent versions
% employed for our CMBfast chains, and is therefore not a limitation of
% DASh itself (Knox 2003, private communication).
% From here on, we use CMBfast % 4.4
% % MT: THAT'S NOT TRUE: LICIA USED 4.2 and I USED  4.3, NO?
% to calculate all our
% likelihood samples.

%\begin{multicols}{2}

The mean scatter in the values of $\ln \Ell$ range from a good
$0.05$ for the 6th order fit to the $\Lambda$CDM 6 parameter model
to a more dubious $0.69$ for the 6 parameter + tensor perturbation
case. Accuracies for all the cases are shown in table
\ref{resultstable}. However it is not immediately clear that the
error in $\ln \Ell$ is necessarily the most interesting quantity,
as it may include contributions at low $\ln\Ell$ which will have
negligible impact on the relevant parts of the likelihood surface.

\begin{figure*}[ht]
%\begin{tabular}{ll}
% {\epsfxsize=4.0cm\epsffile[18 144 592 1018]{devfromgauss.ps}} & {\epsfxsize=4.0cm\epsffile[60 100 450 200]{testpaccuracy.ps}}\\
% \centerline{\epsfxsize=4.0cm\epsffile{testaccuracy.ps}}  & \\
%\end{tabular}
\centerline{\epsfxsize=9.0cm\epsffile{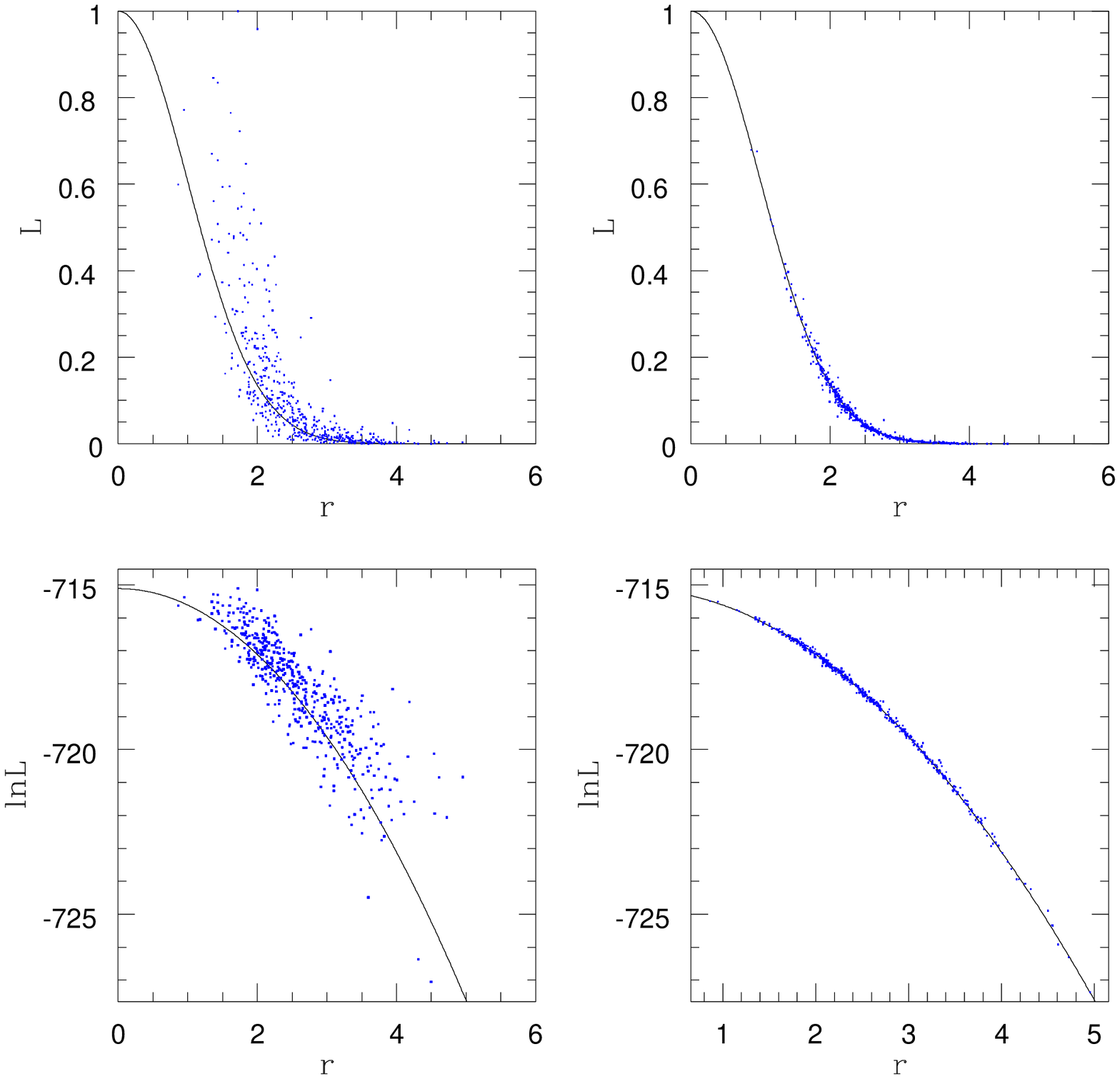}} \caption{The
accuracy of our method for the case of a 5th order polynomial fit
to the 7-parameter case including $\alpha$ is shown for the
likelihood $\Ell$ (top) and its logarithm $\ln\Ell$ (bottom). The
left panels show how well the likelihood is fit by a pure
Gaussian, as a function of the radius $r$ in the transformed
parameter space. The right panels show the corresponding errors
$\Delta\Ell\equiv\Ell_{fit} - \Ell$ (top) and
$\Delta\ln\Ell\equiv\ln\Ell_{fit} - \ln\Ell$ for our polynomial
fit, plotted relative to the Gaussian, \protect\ie, the top right
panel shows $\Ell_{gauss}+\Ell_{fit}-\Ell$. } \label{accfig}
\end{figure*}

An equally interesting quantity is $\Delta \Ell = \Ell_{fit} -
\Ell$ (we normalize $\Ell$ to equal unity at its maximum), which
shrinks rapidly with decreasing values for $\ln \Ell$ and thus
better illustrates what kind of accuracies we can expect for the
marginalized distributions. The mean scatter in the values of
$\Ell$ are significantly smaller, typically of order $\sim 0.01$
for the entire dataset. A better understanding of these quantities
can be obtained by studying figure \ref{accfig}. It shows first
how well (or rather how badly) the actual likelihood surface (both
in terms of $\Ell$ and $\ln \Ell$) is described by a Gaussian PDF
in the transformed normal parameters. These are the variables $\z$
defined in section \ref{fitting}, which have zero mean and
identity covariance matrix, and the $d$-dimensional radius is
given by $r = |\z|=(z_1^2 + \cdots + z_d^2)^{1/2}$. The plot then
shows the fitting errors $\Delta \Ell \equiv \Ell_{fit} - \Ell$
and $\Delta \ln \Ell = \ln \Ell_{fit} - \ln \Ell$, plotted as
errors relative to a Gaussian distribution. This visualizes the
ability of the method to reproduce the likelihood surface with
good precision and the dramatic improvement gained from going to
polynomial degree higher than two.

%\end{multicols}

\begin{center}
\begin{table*}
\begin{tabular}{|c|c|c|c|c|c|c|}
    \hline
 %    & & & & & & \\
    Model & Chain origin & Chain length & Priors &Pol.~deg. $n$ & $\Delta\ln\Ell_{train}$ & $\Delta\ln\Ell_{test}$ \\ \hline
     &  &  &  &  &  &  \\
  6 par T & Own & 183008 & None &6 & 0.02 & 0.08   \\
  6 par T+X & Own & 311391 & None &6 & 0.02 & 0.05   \\
  6 par + $k$ & WMAP & 278325 & $\tau < 0.3$ & 4 & 0.25 & 0.36   \\
  6 par + $\alpha$ & WMAP & 435992 & $\tau < 0.3$ & 4/5 & 0.12/0.08 & 0.14/0.12    \\
%  6 par + $\nu$ & Own & 133361 & & 4 & 0.48  & 0.61   \\
  6 par + $r$ & Own & 178670 & None &4 & 0.29 & 0.69   \\
%  6 par + $w$ & Own &  &  &  &  &  \\
   &  &  &  &  &  &  \\
\hline
\end{tabular}
\caption{R.m.s.~polynomial fit errors for all the
considered cases, ranging from 6 parameter vanilla $\Lambda$CDM models to
spiced up models including curvature, tensor perturbations,
running tilt.
% or massive neutrinos.
} \label{resultstable}
\end{table*}
\end{center}
%\begin{multicols}{2}
%\end{widetext}

\subsection{Application to MCMC}
\begin{figure*}
\centerline{\epsfxsize=10.0cm\epsffile{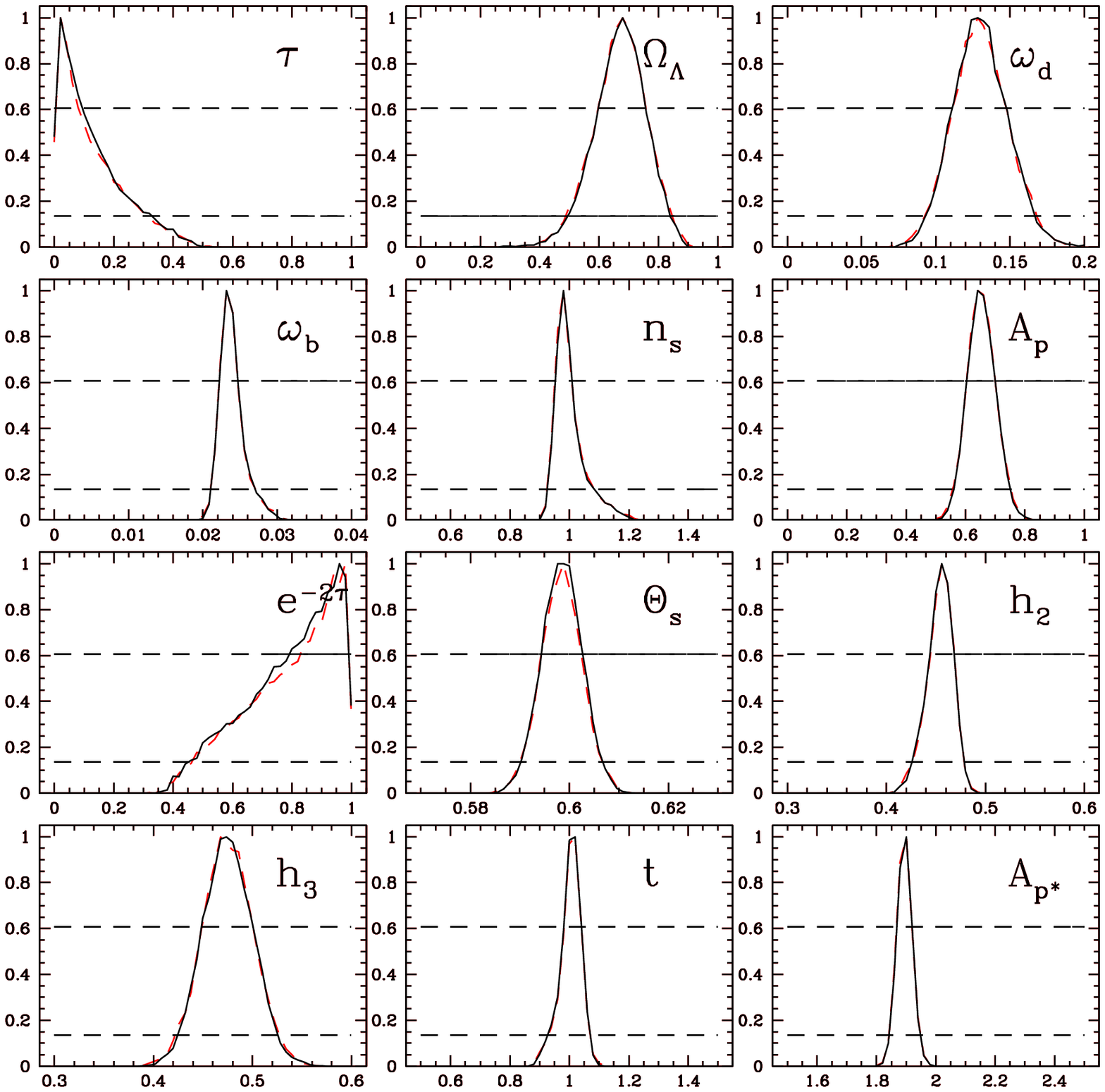}}
\caption{WMAP 6 parameter $\Lambda$CDM case, using only the WMAP
temperature power spectrum. We show marginalized likelihoods for
the polynomial fit compared to the original chain. The two chains
are indistinguishable.} \label{wmap6par}
\end{figure*}

\begin{figure*}
\centerline{\epsfxsize=10.0cm\epsffile{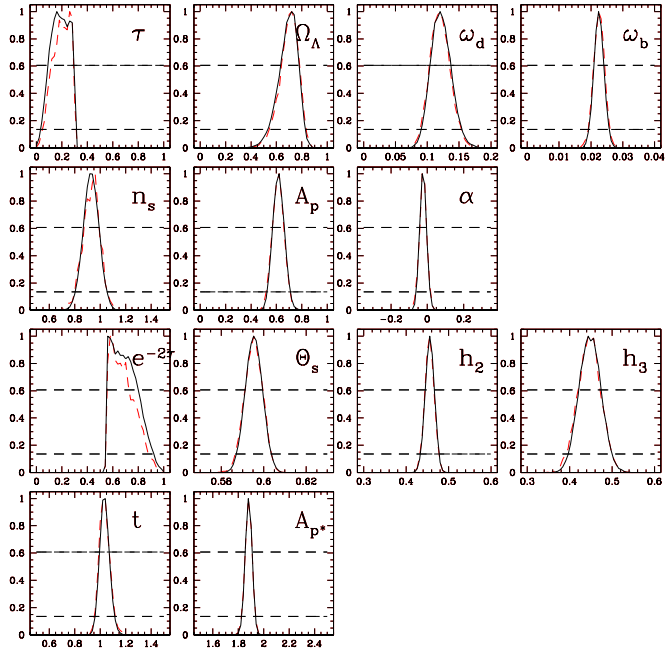}}
\caption{WMAP $\Lambda$CDM + running scalar tilt, marginalized
likelihoods for the polynomial fit compared to the original chain.
Apart from minor differences in the $\tau$ distribution, the two
distributions are identical.}\label{wmappollicia7paralfig}
\end{figure*}

\begin{figure*}
\centerline{\epsfxsize=10.0cm\epsffile{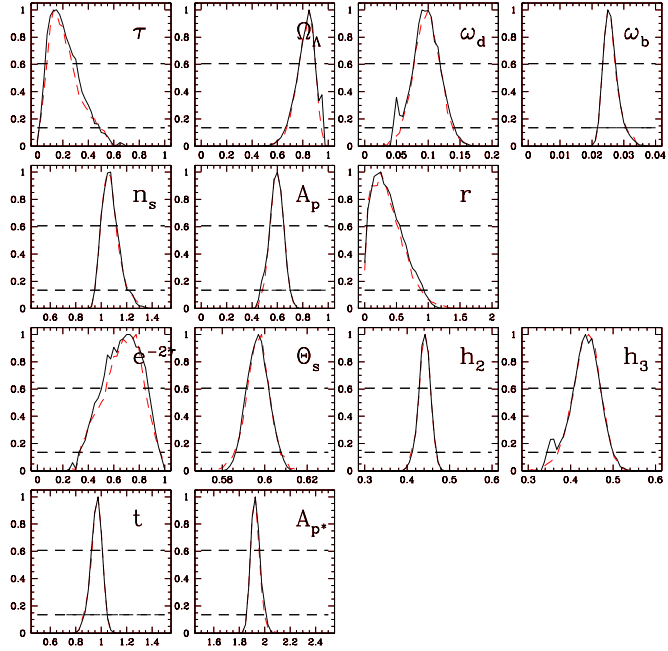}}
\caption{WMAP $\Lambda$CDM + tensor case, marginalized likelihoods
for the polynomial fit plotted over the ones from the original
chain. The difficulty in fitting this case is seen in the
unphysical polynomial artifacts shown in the distributions for
$\omega_d$ and $h_3$. This effect is removed when adding the SDSS
data.} \label{wmappol7parr}
\end{figure*}

The ultimate end-to-end test of the method is how well the
1-dimensional and 2-dimensional marginal distributions are reproduced. To test this,
we use the polynomial fits to calculate the likelihoods for our
MCMC algorithm. Due to the inevitability of polynomial artifacts
at low confidence areas, we use a cut-off at the 3 sigma level
(corresponding to a maximum value for $r$ defined above), where we
replace the polynomial fit by a simple Gaussian. This allows for
the algorithm to find its way through the burn-in process to the allowed part of parameter
space from any starting point.
The results are shown in figures \ref{wmap6par}-\ref{wmappol7parr}.
The figures show what is already indicated in
table \ref{resultstable}: The marginalized distributions are
reproduced to well within the sampling errors, the only exceptions
being a couple of parameters in the tensor case,
% and neutrino cases
demanding only negligible computational time
(a chain of length 200000 runs in about a minute).

\begin{figure*}
\centerline{\epsfxsize=10.0cm\epsffile{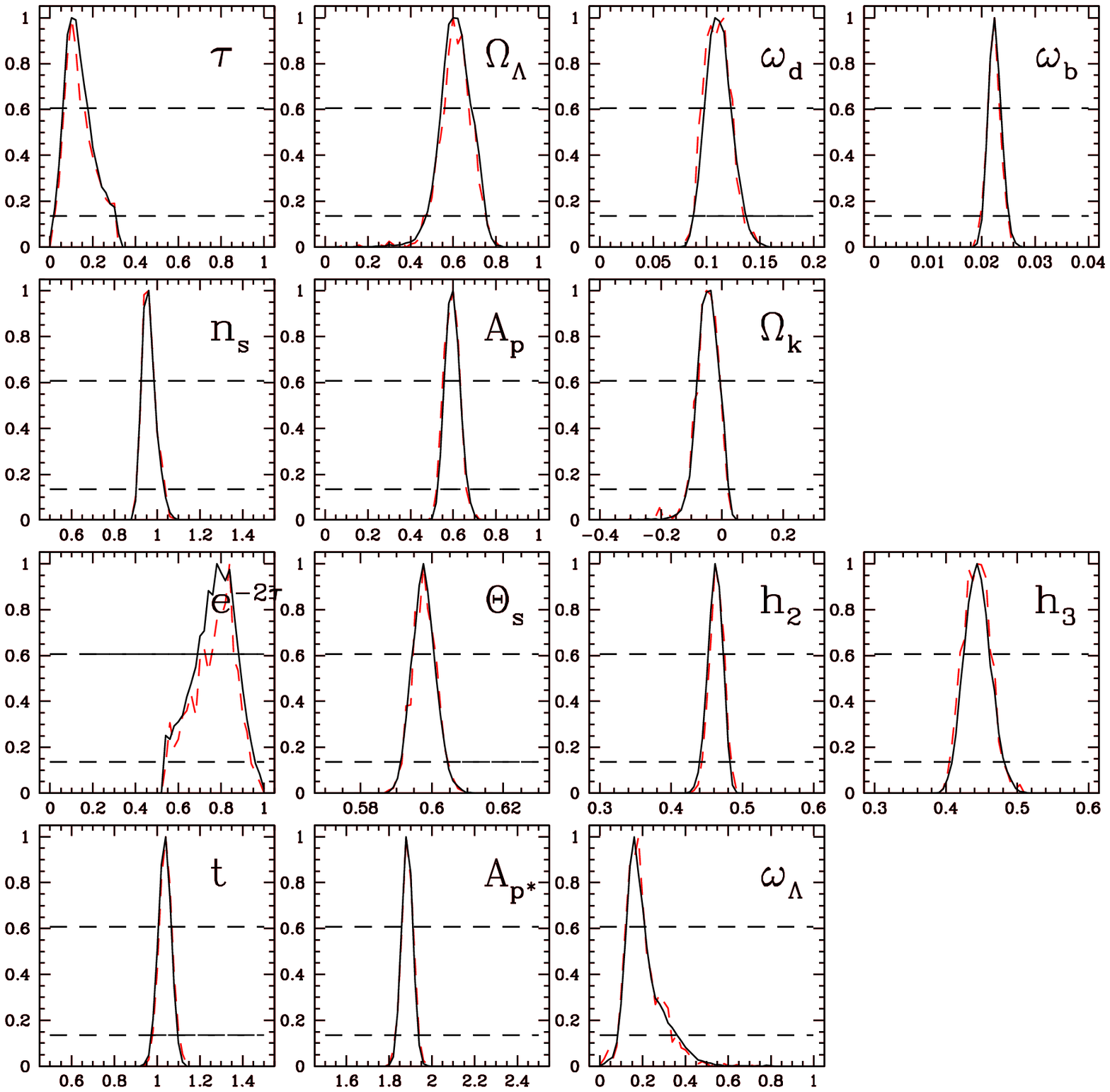}}
\caption{SDSS + WMAP $\Lambda$CDM + curvature contributions,
marginalized likelihoods for the polynomial fit compared to a
chain consisting of only $\sim 14000$ points. The fit is
excellent, the errors coming from poisson noise due to the short
length of the CMBfast calculated chain.} \label{sdsswmappol7par}
\end{figure*}

Figure (\ref{sdsswmappol7par}) illustrates one of the most useful
applications of our approach: combining WMAP with another data
set, here the SDSS galaxy power spectrum as reported in
\cite{SDSS,SDSS2} to give constraints on the cosmological
parameters for a non-flat 7 parameter model. The red dashed curve
shows the results from running a Markov Chain Monte Carlo for
about a week obtaining only 14000 points, and the solid black
curve is the reproduction by our polynomial fit, taking a mere
afternoon to run. Indeed, the time-consuming part in this
calculation was the computation of the non-linear matter power
spectrum, the processor time needed to calculate the WMAP
likelihoods being under a minute.

%\subsection{$\Lambda$CDM 6 parameter model}
%This section presents the results of our work for 6 parameter
%$\Lambda$CDM models using TT only and TT+TE power spectra.
%\end{multicols}

\begin{figure*}
%\centerline{\epsfxsize=10.0cm\epsffile{wmappol6parfit1dcomp.ps}}
\centerline{\epsfxsize=10.0cm\epsffile{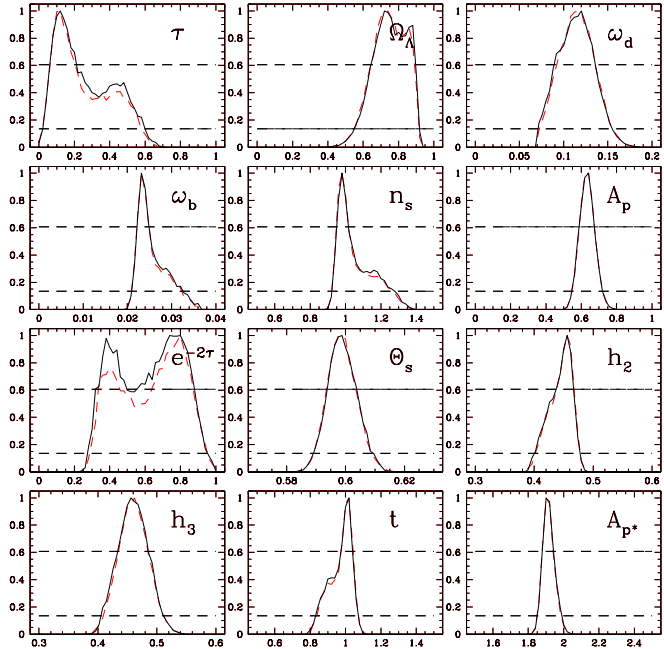}}
\caption{WMAP $\Lambda$CDM T+X case, marginalized likelihoods for
the polynomial fit plotted on top of the ones from the original
chain. The two chains are nearly indistinguishable apart from a
small difference in fitting the second (and otherwise ruled out)
peak in the optical depth distribution.}
\end{figure*}

%\subsection{$\Lambda$CDM + 7th parameter}
%Here we show the results for 7 parameter ``spiced up" $\Lambda$CDM
%models using TT+TE power spectra, including non-flat universes,
%massive neutrino-contributions, tensor perturbation contribution
%as well as running tilt and a dynamical dark energy scenario $w
%\neq -1$.

\begin{figure*}
\centerline{\epsfxsize=10.0cm\epsffile{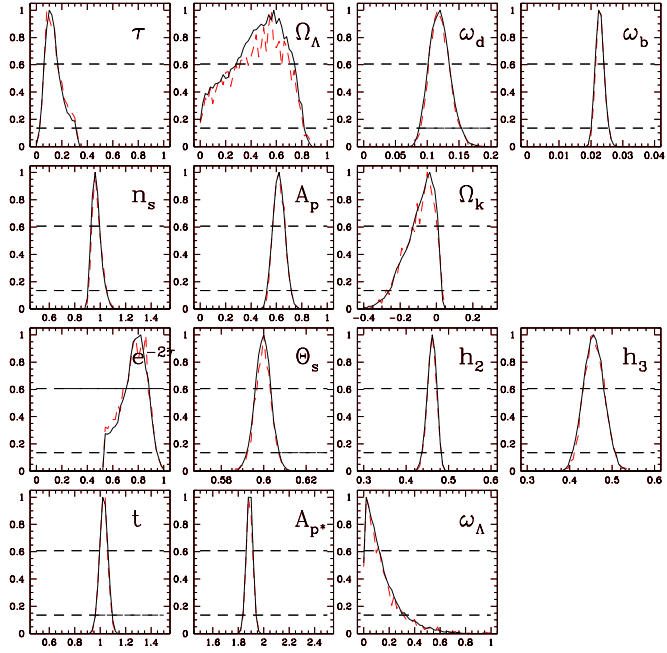}}
%\centerline{\includegraphics{wmappol7parfit1dcomp_wl_ownchain.gif}}
\caption{WMAP $\Lambda$CDM + curvature case, marginalized
likelihoods for the polynomial fit compared to the original chain.
Again we see two almost identical sets of distributions, apart
from some poisson noise in the original chain.}
\end{figure*}

%\begin{figure}
%\centerline{\epsfxsize=12.0cm\epsffile{wmappol7parnufit1dcomp.ps}}
%\caption{WMAP $\Lambda$CDM + massive neutrino contributions,
%marginalized likelihoods for the polynomial fit (yellow/light
%grey) compared to the original chain (red/dark grey)}
%\end{figure}
%\subsection{$\Lambda$CDM + Running Tilt; TT+TE likelihoods}
%\subsection{$\Lambda$CDM + Tensor Perturbation Contributions;
% TT+TE likelihoods}
%\subsection{$\Lambda$CDM + Dark Enegy equation of state $w \neq
% -1$}
%\begin{multicols}{2}

\section{Conclusions and Discussion}

The big bottleneck in parameter estimation with CMB data has
hitherto been the computation of theoretical power spectra by
means of CMBfast, CAMB, CMBEASY, DASh or similar, for a reasonably
fast computer typically requiring from $~5$ seconds to several
minutes for non-flat models with massive neutrino for CMBfast, and
$~1$ second for DASh. In contrast, our method requires less than a
millisecond per model, since we have already precomputed the WMAP
likelihoods using CMBfast and distilled the results into a
convenient fitting function. After transforming into a physically
motivated parameter set, the only calculation needed is the
evaluation of an $n^{th}$ order polynomial with as few as 210
coefficients for the 6 parameter quartic case. As the number of
coefficients may actually exceed the number of measured $C_l$s the
method is clearly not a way of compressing the data. Rather we are
compressing the amount of calculation necessary to convert
cosmological parameters into likelihood values.

The typical error in this approach is $\Delta\Ell\sim 0.01$ (with
the peak likelihood normalized to unity) and $\Delta\ln\Ell\sim
0.05 - 0.5$. To place these inaccuracies in context, let us first
discuss how they
%Since it
%is unclear which of these numbers are the most relevant the
compare with the inaccuracies in other methods, then  the question of how good is good enough.

%The error in this approach is of order $\lesssim 0.1\%$, which is
%tiny compared to the $1-2\%$ inaccuracy inherent in CMBfast.
%\subsection{State of the art accuracies}

The latest version of CMBfast\cite{seljak03} has an accuracy down
to $\sim 0.1\%$ (r.m.s). It is unclear if other recent packages
match this number, but they are generally good to the $1-2\%$
level. Despite computer advances, the CMBfast code is still slow
when several hundred thousand model computations are required as
for grid calculations or Markov Chain Monte Carlo applications.
Several shortcuts have therefore been developed in order to
compute the Likelihood faster than the full CMBfast.
% and CAMB codes  MT: see Matias' comment about about CAMB being essentially CMBfast
The ``$k$-split" method in CMBfast
due to Tegmark {\etal} \cite{ksplit} utilizes the fact that the high-$\l$
and low-$\l$ ends of the power spectrum depends on different
parameters, and so calculates the two parts separately. The end
result is then combined into a final power spectrum. The Davis
Anisotropy Shortcut (DASh)\cite{dash} takes this method further,
and creates the power spectrum by interpolation between points in
a huge pre-computed grid of transfer functions.
% MT: PERHAPS WE SHOOT OURSELVES IN THE FOOT IF WE SAY THIS, GIVEN THE NUMBER AND
%     THE DISCUSSION I ADDED BELOW.
% These methods have
% now been around for a while and their accuracies were good
% enough given the observational data available at the time of
% publishing. It may be time to test these accuracies again now that
% WMAP combined with other data offers accuracies of order a few per cent
% in cosmological parameter estimation\cite{eisenstein}.
Comparing likelihoods calculated with $k$-split CMBfast and DASh
with those calculated with the maximally accurate CMBfast, we found that the errors from our
fitting method in both $\Delta\ln\Ell$ and $\Delta\Ell$ are
significantly smaller than the errors in these quantities from
the DASh
%\footnote{We emphasize that our version of
%DASh used a grid calculated by CMBfast version 3.2 and the errors
%appear to be dominated by this rather than by inherent DASh
%inaccuracies.}
or $k$-split approximations.
For instance, DASh gave r.m.s.~$\Delta\ln\Ell\approx 0.9$ for the 6-parameter
case using only unpolarized WMAP information and the ksplit inaccuracies are similar, which should
be compared with our value $\Delta\ln\Ell\approx 0.08$ from Table II.
% MT: I got dash rms dlnL = 0.86 for wmap6par test

How accurate is accurate enough?
For most applications, the key issue is not inaccuracies in
the power spectra or likelihoods, but inaccuracies in measured cosmological
parameters. If an approximation shifts the measured values of all parameters
by much less than their statistical uncertainties, it is for all practical
purposes irrelevant.

We have also shown that the errors in the marginalized
distributions are minimal, and that our method should be used by
anyone who do not have excessive amounts of time and computer
power on their hands. Our last six figures show that our
approximation is clearly accurate enough in this sense except for
the above-mentioned glitches in the tensor case. There is,
however, one subtle and somewhat counterintuitive point that is
worth clarifying. Since WMAP constrains the power spectrum
normalization to the $10^{-3}$ level when all other parameters are
held fixed, this means that even a seemingly tiny $\sim 0.1\%$
inaccuracy in the power spectrum can in principle cause a change
of order unity in $\chi^2$ and $\Delta\ln \Ell$, \ie, an
inaccuracy larger than that of our fitting method, and one may
naively expect that inaccuracies of order $\Delta\ln \Ell\sim 1$
would affect the parameter measurements at the $1\sigma$ level.
Although this would be true if only one parameter were being
measured, the situation at hand is actually much better because of
degeneracies. As long as the inaccuracies do not exactly mimic the
effect of changing some cosmological parameter, they will alter
$\ln \Ell$ mainly via the narrowest directions in the
multidimensional degeneracy banana and hence have little effect on
the final results. For instance, if a fitting inaccuracy causes a
relative error in $\Theta^E_s$ at the $10^{-3}$ level, it will
have no noticeable effect on the estimates any of the vanilla
cosmological parameters, whose error bars are dominated by the
eigenparameters with the largest rather than the smallest
uncertainties. The bottom line is therefore that although the the
DASh and $k$-split $\ln \Ell$ inaccuracies of order unity may
sound worrisome, these two approximations are nonetheless
sufficient to give negligible inaccuracies in cosmological
parameter measurements, and the still better accuracy of our
fitting method is actually overkill.

Our method has several applications. First, it is
extremely useful when combining the WMAP-data with non-CMB data
such as the galaxy surveys, weak lensing data, supernovae data,
{\etc}, where likelihood calculations are fast relative to CMB power spectrum
calculations. Using our method to measure parameters from the combined
WMAP and SDSS data \cite{SDSS2} enabled us to do some of the analysis
with dramatically increased speed\footnote{
The analysis in \cite{SDSS2} used no chains from the WMAP team,
but applied our fitting method to
the WMAP Monte Carlo Markov Chains described in Table 8 of that paper.}.
Second, even when
running Markov chains for models not considered here, one can get
good results by simply running the chain for long enough to
acquire a sufficient (but not necessarily statistically fair)
sample of the surface, then compute the polynomial fit, and go on
to use this fit for the remainder of the job --- until all relevant
mixing and convergence tests are fulfilled.
%This approach can be further improved
%by thinning the resulting polynomial fit chain by a factor say 100 to make
%all points uncorrelated and then importance sampling as described by \cite{cosmomc}
%with CMBfast to get the exactly correct distribution.
A further obvious application is generating extremely long
Markov chains, eliminating sampling errors almost completely.
Finally, should the reader be intent on requiring the best accuracy
that CMBfast can offer, our approach can still be helpful: It is
a fundamental fact about MCMC algorithms that they do not produce
completely uncorrelated points, and the correlation length of the
chains can easily be several hundred points,
and only gets as low as 45 even for our simply 6-parameter WMAP chains.
Thus the calculations may be
significantly accelerated by creating a statistically random sample by means
of the polynomial fit, thinning the chain to every 200 points or
so, and calculating the CMBfast power spectrum and WMAP
likelihoods for these points. A final likelihood sample with the strictly correct
distribution may then
be obtained using importance sampling as described in \cite{cosmomc}.

We supply Fortran routines for computing the likelihoods for all
the cases given in the text at
http://www.hep.upenn.edu/{$\sim$}sandvik/CMBfit.html , and we plan
to complement this work with further models and fits in the
future.

\section{Acknowledgements}
We wish to thank L. Verde and the rest of the WMAP team for kindly
supplying us with the Markov chains from their analysis. HS wishes
to thank R.~Jimenez and P.~Protopapas for valuable discussions.
This work was supported by NSF grant AST-0134999, NASA grant
NAG5-11099 and fellowships from the David and Lucile Packard
Foundation and the Cottrell Foundation.

\end{multicols}

\end{document}